\documentclass[]{spie}  

\usepackage{amsmath,amsfonts,amssymb}
\usepackage{graphicx}
\usepackage{upgreek}
\usepackage[colorlinks=true, allcolors=blue]{hyperref}
\usepackage{acronym}

\acrodef{AO}[AO]{adaptive optics}
\acrodef{EE}[EE]{encircled energy}
\acrodef{ELT}[ELT]{Extremely Large Telescopes}
\acrodef{HR}[HR]{hybrid reformatter}
\acrodef{MFD}[MFD]{mode field diameter}
\acrodef{MM}[MM]{multi-mode}
\acrodef{NIR}[NIR]{near infrared}
\acrodef{SM}[SM]{single-mode}
\acrodef{PD}[PD]{photonic dicer}
\acrodef{PL}[PL]{photonic lantern}
\acrodef{PSF}[PSF]{point spread function}
\acrodef{ULI}[ULI]{ultrafast laser inscription}

\title{Optimizing astrophotonic spatial reformatters using simulated on-sky performance}

\author[a,e] {Theodoros Anagnos} 
\author[a] {Robert J. Harris}
\author[b] {Mark K. Corrigan}
\author[c] {Andrew P. Reeves}
\author[b] {Matthew J. Townson}
\author[d] {David G. MacLachlan}
\author[d] {Robert R. Thomson}
\author[b] {Tim J. Morris}
\author[e,f] {Christian Schwab}
\author[a] {Andreas Quirrenbach}

\affil[a]{Landessternwarte, Zentrum f\"ur Astronomie der Universit\"at Heidelberg, K\"onigstuhl 12, 69117
Heidelberg, Germany}
\affil[b]{Centre for Advanced Instrumentation, Durham University, South Road, Durham DH1 3LE, UK}
\affil[c]{Deutsches Zentrum f\"ur Luft-und Raumfahrt (DLR), Oberpfaffenhofen, 82234 We\ss ling, Germany}
\affil[d]{SUPA (Scottish Universities Physics Alliance), Institute of Photonics and Quantum Sciences, Heriot-Watt University, Edinburgh, EH14 4AS, UK}
\affil[e]{Department of Physics and Astronomy,
Macquarie University, NSW 2109, Australia}
\affil[f]{The Australian Astronomical Observatory (AAO), Level 1, 105 Delhi Rd, North Ryde, NSW, 2113, Australia}

\authorinfo{Further author information: (Send correspondence to Th.A.)\\Th.A.: E-mail: tanagnos@lsw.uni-heidelberg.de, Telephone: (+49) 06221-54-1727}

\begin{document} 
\maketitle

\begin{abstract}
One of the most useful techniques in astronomical
instrumentation is image slicing. It enables a
spectrograph to have a more compact angular slit,
whilst retaining throughput and increasing resolving
power. Astrophotonic components like the photonic
lanterns and photonic reformatters can be used to
replace bulk optics used so far. This study investigates 
the performance of such devices using end-to-end
simulations to approximate realistic on-sky 
conditions. It investigates existing components,
tries to optimize their performance and aims to
understand better how best to design instruments
to maximize their performance. This work complements
the recent work in the field and provides an 
estimation for the performance of the new components.
\end{abstract}


\section{INTRODUCTION}
\label{sec:intro}
Over last decade one of the major goals driving the
development of astronomical instrumentation has been
the detection of an Earth-like planet orbiting either
a Sun-like star or an M-dwarf. Whilst this is 
challenging, the science driver is also very exciting;
the exploration of small possibly-habitable exoplanets
inside the Goldilocks zone (e.g., Ref. \citenum{Mayor:2003,
Quirrenbach:2016}). One of the most successful methods
to date is the radial velocity technique, where the 
presence of a planet is inferred through movements
in the stellar lines. To use the radial velocity 
method to detect these small planets, sub-m/s radial
velocity precision in the measurements is needed. 
This requires a highly stable spectrograph that has
been carefully calibrated. 

Harnessing the temporal stability of the \ac{SM} 
fiber's spatial profile output (close to the 
diffraction limit) as the spectrograph's input, is
one of the solutions to making this task easier 
(e.g., Ref. \citenum{Coude:1994,Crepp:2014,Schwab:2014,
Jovanovic:2016}). However, the atmosphere limits the
coherence of the light, forming a seeing disk instead
of a diffraction limited \ac{PSF} resulting in 
considerable coupling losses. As a consequence, most
current astronomical telescopes operate in the seeing
limited regime - the \ac{MM} regime of photonics,
where coupling of starlight is easier than in the
\ac{SM} regime. However, the required size of the
spectrograph gets larger in the seeing limited regime,
demanding large fragile optics in order to maintain
high spectral resolving power (R $>$ 100,000) on
large telescopes \cite{Bland-Hawthorn:2006}.

In the seeing limited regime, the size of a dispersive
spectrograph is dependent on the diameter of the
telescope \cite{Anagnos:2018}. This dependence imposes
a correlation of the number of the spatial modes 
forming the telescope's \ac{PSF} being analogous to
the ratio of square of the telescope's diameter
$D_\mathrm{T}$ over the Fried seeing parameter $\mathrm{r_{0}}$ 
\cite{Harris:2013,Spaleniak:2013,MacLachlan:2017}.

Currently the largest ground-based telescopes are
8-10 m in diameter requiring large spectrographs,
to couple the light efficiently while also achieving
high resolving power (e.g., Ref. \citenum{Vogt:1994,
Noguchi:2002,Tollestrup:2012}). This problem will be
even more challenging for spectrograph designers
when the \acl{ELT} (ELT, GMT, TMT) are built
\cite{Cunningham:2009,Mueller:2014,Zerbi:2014}.

A solution for this is to use \ac{AO} systems to
reduce the number of spatial modes. Some extreme
\ac{AO} systems manage to get close to the diffraction
limited regime ($>$ 90 \% Strehl ratio) in the H-band,
but these demand a very bright guide star to operate
properly (e.g., Ref. \citenum{Dekany:2013,Agapito:2014,
Macintosh:2014,Jovanovic:2015}). Their performance
also degrades at visible wavelengths, limiting the 
available science targets.

Another approach, with less severe constraints
on image quality is to reduce the size of the
instrument by spatial reformatting. By using this
technique, the \ac{PSF} is  reformatted into a 
different geometry. This is conceptually similar to
image slicing (Ref. \citenum{Weitzel:1996},
and references therein). The sampled \ac{PSF} can be
divided in smaller pieces that can be coupled to 
smaller and more stable instruments
\cite{Allington-Smith:2004,Hook:2004,Harris:2013}.
It should be noted, however, that most implementations
of reformatting do not retain spatial information 
and therefore are not suitable for scientific goals
requiring spatially resolved information.

Harnessing photonic technologies to implement this
concept has led to a variety of new devices, to 
name a few the PIMMS (Photonic Integrated Multimode
Micro Spectrograph) \cite{Bland-Hawthorn:2010}, the
\ac{PD}\cite{Harris:2015}, the \ac{HR}\cite{MacLachlan:2017}
and the Photonic TIGER device\cite{Leon-Saval:2012}.
The majority of these devices are derived from the
\ac{PL}\cite{Leon-Saval:2005,Leon-Saval:2013,Birks:2015},
and consist of different optical fiber/inscribed 
waveguides geometries. The \ac{PL} is a device that
bridges the \ac{MM} and the \ac{SM} regimes by
having a \ac{MM} core to the one end and many \ac{SM}
cores at the other end. Initially, \acp{PL} were
constructed with fibers
(e.g., Ref. \citenum{Birks:2015,Yerolatsitis:2017}),
but later were manufactured using other methods \\
(e.g., Ref. \citenum{Thomson:2011,Spaleniak:2013}).
They allow the \ac{PSF} to be coupled into
instruments more efficiently than \ac{SM} fibers
\cite{Cvetojevic:2009,Cvetojevic:2012}.

A major benefit of operating in the \ac{SM} regime
is the absence of modal noise in spectroscopic
measurements, enabling better calibration of the 
acquired measurements than before\cite{Probst:2015}.
Coupling the time changing \ac{MM} input to the
spectrograph creates the modal noise, which induces
movement of the barycenter for any given wavelength.
As a result, more noise is added into the 
spectroscopic measurements reducing the precision
(e.g., Ref. \citenum{Lemke:2011,Perruchot:2011,
McCoy:2012,Bouchy:2013,Iuzzolino:2014,Halverson:2015}).
Conversely, by using a \ac{SM} fiber the modal noise
is eliminated, as only the fundamental mode propagates
(not counting different polarization states), while
the higher order modes radiate out into the cladding
of the fiber.

Over the last decade, astrophotonic reformatters
have been used to combine the high coupling
efficiency of a \ac{MM} fiber and the noise
elimination properties of a \ac{SM} fiber. However,
special care should be given to the instrument 
configuration for sources of modal noise at early 
stages of the optical set-up\cite{Spaleniak:2016,
Cvetojevic:2017}. Furthermore, there is no spatial
preservation of the coupled image into astrophotonic
reformatters in comparison to a conventional 
image slicer. Nevertheless, this is not a limitation
for high resolution spectroscopy, as these devices
are often placed after a \ac{MM} fiber and
additional components such as scramblers and modal
noise elimination devices, producing a stable
\ac{PSF} (Ref. \citenum{Quirrenbach:2016}),
but not preserving the spatial information.

In this paper, we analyze the simulated results of
an astrophotonic reformatter, namely the \ac{HR}
\cite{MacLachlan:2017}, and compare it with its 
on-sky performance. This is an astrophotonic
spatial reformatter that geometrically manipulates
the \ac{MM} input \ac{PSF} into a \ac{SM} 
(diffraction-limited) pseudo-slit output aiming to
increase the resolving power of the spectrograph
into which the light is coupled, and enable more
precise measurements of astronomical targets with the
elimination of modal noise. The \ac{HR} is composed
of a 92-core multicore fiber (91 plus an extra core
to define orientation and aid alignment) tapered 
down to form a \ac{PL}\cite{Leon-Saval:2005,
Leon-Saval:2013,Birks:2015} having a \ac{MM} 
entrance at one end face, while towards the other
end the cores are uncoupled in the original 
hexagonal arrangement of the multicore fiber. 
Following that, the cores are connected to a 
\ac{ULI} reformatter that re-arranges the 
hexagonal geometry of the multicore fiber cores
into a slit profile output to feed a spectrograph.

This paper is organized as follows: starting with
Section \ref{sec:methods} the configuration 
parameters for the simulation of the \ac{HR} are
reported. In Section \ref{sec:results}
results are presented as well as the selected method
and the optimization techniques. Following that, a
discussion of the results is included in Section
\ref{sec:discussion}, and finally we conclude in\\
Section \ref{sec:conclusion}.

\section{Methods}
\label{sec:methods}

Performing precise simulations is of paramount
importance for calibration of future designs as 
well as to estimate their performance in realistic
conditions. Motivated by this, we used a combination
of two simulation tools to simulate an astrophotonic
reformatter, namely the \ac{HR}\cite{MacLachlan:2017}.
We made use of: Soapy \cite{Reeves:2016}, a Monte 
Carlo \ac{AO} simulation tool to replicate the
atmospheric conditions encountered on-sky and the
influence on the device's performance; and BeamProp
by RSoft \cite{rsoft}, which is a finite-difference
beam propagation solver to model the light propagation
through the components.

The configuration of the simulation was the following:
first, Soapy was configured to produce an \ac{AO} - 
corrected star, and then the frames were used as an
input for the BeamProp simulations.

\subsection{Soapy set-up}
The \ac{HR} was tested on-sky with CANARY\cite{Myers:2008}
at the William Herschel Telescope. Therefore, Soapy
was configured to approximate the sky conditions
during the observation run (see Table \ref{soapy_set}).
Soapy was used to simulate three different \ac{AO}
operation modes as on-sky with CANARY, namely closed-loop, 
open-loop and tip-tilt. In closed-loop \ac{AO} mode of
operation CANARY provided correction at an update
rate of 150 Hz, both for tip-tilt and higher order
aberrations, while for the tip-tilt \ac{AO} mode of
operation the integrator loop gain on the high-order
modes was reduced to a small value of the order of 
0.001, resulting in the correction of the \ac{PSF}
position in real-time, but not for the \ac{PSF} 
shape which was provided without high-order \ac{AO}
correction. As for the case of open-loop mode of
correction, the gain on the tip-tilt correction
was further reduced in order to have the \ac{PSF}
remain in the reference location providing optimum
coupling, but without having high-temporal frequency
correction. Also, a value of 0.15 m for the seeing 
parameter (Fried parameter-$\mathrm{r_{0}}$) was
selected, representative of the aforementioned
on-sky experiments\cite{MacLachlan:2017}.

Soapy produced 200 \ac{NIR} data frames, each with
a 6 ms exposure time as an input to BeamProp. Each
Soapy frame covers a 3.0 arcseconds squared field 
on-sky, slightly more than three times the \ac{HR}
entrance angular size. In contrast to the data 
acquired on-sky, the Soapy frames provided  both 
phase and amplitude information, which proved to 
be crucial for the simulation results (see Section
\ref{sec:throughput}).

\begin{table*}
 \caption{Soapy input parameters for simulation}
 \label{soapy_set}
 \centering
 \begin{tabular}{l c c c}
 \hline\hline
  & & Modes of \ac{AO} operation\\
 \hline
 & closed-loop & open-loop & tip-tilt\\
 \hline \hline
 Parameters \\
 \hline
   Seeing (arcsec) & 0.69 & 0.69 & 0.69\\
   Instantaneous Strehl ratio (mean) & 0.53 & 0.21 & 0.11\\
   Long exposure Strehl ratio (mean) & 0.38 & 0.04 & 0.06\\
   Fried parameter $\mathrm{r_{0}}$ (m) (@1550 nm) & 0.15 & 0.15 & 0.15 \\
   Atmosphere layers & 5 & 5 & 5 \\
   DM integrator loop gain tip-tilt & 0.3 & 0.001 & 0.3\\
   DM integrator loop gain Piezo & 0.3 & 0.001 & 0.001\\
   \hline\hline
   \end{tabular}
    \label{table:soapy}
\end{table*}

\subsection{BeamProp set-up}

\begin{figure}
\centering
\includegraphics[width=0.6\textwidth]{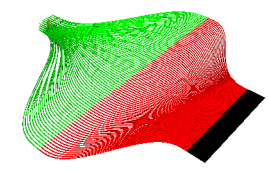}
\caption{The hybrid reformatter 3D structure, without
the initial photonic lantern section included in the
plot, visualized in the RSoft's CAD environment. The
colors show the 3 different transition planes used.}
\label{hybrid:cad}
\end{figure}

In the next stage, the frames from Soapy were used 
as an input for BeamProp simulations; the \ac{HR}'s
angular size was set to 1.1 milliarcseconds. The
design structure of the device was followed as
described in Ref. \citenum{MacLachlan:2017}. The 
\ac{HR} is composed of a \ac{PL} section followed 
by a \ac{ULI} reformatter section, ending in a 
slit profile output.

To perform the simulations in BeamProp, the refractive
index information is a crucial parameter, both for
the core and the cladding materials. For the \ac{HR}
the initial \ac{PL} section of multicore fiber was
made by fused silica having step-index Ge-doped cores 
fabricated as in Ref. \citenum{Birks:2012,Birks:2015},
and for the following reformatter section a substrate
of borosilicate glass was used (SCHOTT AF45)
\cite{Meany:2014} for the inscription of the waveguides
with a $\mathrm{n_{cl}}$ of $\sim$1.4974 at 1550 nm.
As the are no refractive index measurements of the
waveguide profiles for the device, the values
from Ref. \citenum{Thomson:2011} were used
($\mathit{\Delta} = \frac{n_{\mathrm{core}}-n_{\mathrm{clad}}}
{n_{\mathrm{core}}}\approx 1.76\times 10^{-3}$) as
they are considered to be close to our device,
despite the small expected differences resulting 
from the inscription parameters (see Table \ref{table:ULI})
and the difference in glass. The 3D structure of
the \ac{ULI} section of the \ac{HR} as shown in 
RSoft's CAD, is illustrated in Figure \ref{hybrid:cad}.

In general, in the BeamProp simulation tool the material
propagation loss is a freely chosen parameter depending 
on the material and its properties. Therefore, in order
to add greater precision in our simulations we ran tests
using a propagation loss of 0.1 dB/cm\cite{Nasu:2005}.
This value is an optimistic estimate representing 
relatively low losses when compared to the losses due
to the geometric design ($<$ 2 \% over the \ac{HR}
length). Further tests will be performed in future
modeling with optimized devices.

\begin{table}[ht]
 \centering
 \caption{Comparison of ULI inscription parameters
 used in Ref. \citenum{Thomson:2011} and Ref. \citenum{MacLachlan:2017}.}
 \begin{tabular}{lcc}
 \hline
 Parameters & Thomson et al. & \acf{HR}\\
 \hline
 $n_{\mathrm{cl}}$ (@1550 nm) & $\sim$ 1.49 & $\sim$ 1.4974\\
 Pulse Energy (nJ) & 165 & 174\\
 Pulse repetition rate (kHz) & 500 & 500\\
 Pulse duration (fs) & 350 (1047 nm) & 430 (1030 nm)\\
 \hline
 \end{tabular}
 \label{table:ULI}
\end{table}

\subsection{Throughput measurement}
In order to calculate the total throughput 
($T_{\mathrm{tot}}$) and transmission
properties of the device from simulations the 
following formula was used:

\begin{equation}
\centering
\label{eq3}
\hspace{2cm} {T_{\mathrm{tot}} = \frac{F_{\mathrm{out}}
(i)}{F_{\mathrm{in}}(i)},
\quad i = \# frames},
\end{equation}

where $F_{\mathrm{out}}$ is the output of the slit 
and $F_{\mathrm{in}}$ is the sum of the coupled flux
of the input field.

\section{Simulation results}
\label{sec:results}

\subsection{Output power performance results}
\label{sec:throughput}

Here we present the simulation results. BeamProp
simulations were performed using Soapy output data
(amplitude and phase information). As the on-sky
frames obtained with CANARY only record intensity
these were not used with BeamProp. This is due to
the lack of phase information, which yields 
incorrect results\\ (see Ref. \citenum{Anagnos:2018}).

\begin{figure}
\centering
\includegraphics[width=0.6\textwidth]{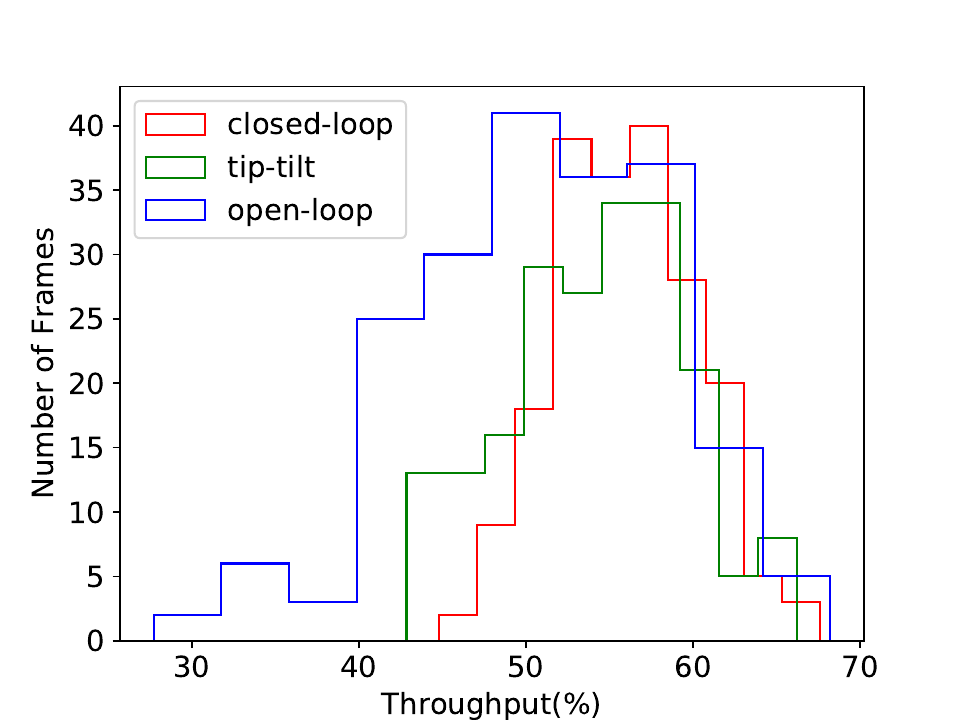}
\caption{Histogram distribution of throughput
measurements from simulations shown for three
different \ac{AO} modes. Red color represents 
the closed-loop \ac{AO} mode of correction,
green shows the tip-tilt correction and blue
shows open-loop correction.}
\label{th_inst}
\end{figure}

Figure \ref{th_inst} shows the results of simulating
200 frames with Soapy. The throughput of the \ac{HR}
for closed-loop \ac{AO} mode of operation was measured
to be 56 $\pm$4 (\%). In tip-tilt operation mode the
output was measured to be 55 $\pm$5 (\%); and for
the open-loop correction results were 51 $\pm$7 (\%).

\begin{figure}
\centering
\includegraphics[width=0.6\textwidth]{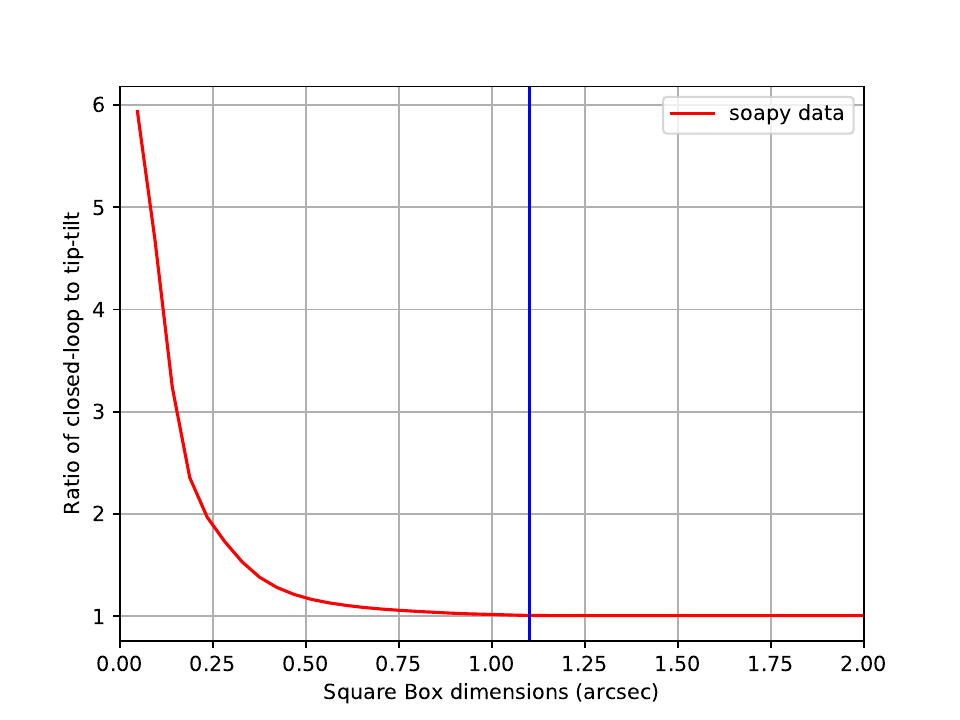}
\caption{The behavior of closed-loop and tip-tilt
\ac{AO} mode ratios of \ac{EE} related to the 
spatial scale (circular radius centered), plotted for
simulated (Soapy) averaged (intensity) data. The
vertical blue line represents the circular entrance
size of the \ac{PL} section of the \ac{HR}.}
\label{ratio_ct}
\end{figure}

The ratio between the closed-loop and tip-tilt
\ac{AO} modes of correction plotted as a function
of the device \ac{MM} entrance input size for the 
Soapy averaged intensity data, is shown in Figure
\ref{ratio_ct}. This is calculated to better
understand the coupling performance. The figure 
shows that the spatial size of the \ac{MM} entrance
of the device is larger than the spatial size of
the input. Hence, all the light is coupled into 
the \ac{HR} because of its large collecting area.

\subsection{Optimization results}

When the \ac{HR} was designed and manufactured in
2014 there was no attempt to fully model it.
Therefore, there are many areas for further 
optimizing the device. To study this, a Monte 
Carlo simulation run was performed to investigate
the optimal length of transition positions in the
\ac{ULI} section of the device (see Figure \ref{hybrid:cad}).

A gradual transition of the \ac{ULI} waveguides position
in between the three different planes (see Figure 
\ref{hybrid:cad}) is helpful to ensure low loss
of light \cite{Birks:2015}. Nevertheless, there is
a trade off between material and bend losses against
the device's length, when making use of the \ac{ULI}
technique to inscribe waveguides in materials.

In order to optimize further the throughput
performance of the \ac{HR}, a Gaussian input 
(close to the diffraction limit) with a 25 $\upmu$m
\ac{MFD} generated by RSoft was chosen. Injecting
this input, we performed a short Monte Carlo 
simulation on the \ac{HR}, in order to optimize
each of its reformatter section transition planes
for transmission by scanning for a variety of 
different lengths between the 3 transition planes
of the reformatter section. Preliminary results
show an improvement in theoretical transmission
by 2\% (a $\sim$ 6\% improvement in performance),
leading to a shorter more compact device. However,
transition planes can be further improved to be
lower in loss (more adiabatic).

\subsection{Modal noise results}
\label{sec:mnoise}
Aiming to investigate the modal noise properties
of our theoretical \ac{HR} device, we followed
a similar procedure to that described by Ref.
\citenum{Anagnos:2018}. We choose the 1550 nm 
wavelength region to perform our simulations. We
analyze the results from it by measuring the
barycentric movement of the slit output 
\cite{Rawson:1980,Chen:2006} examining its 
stability in the near field regime.

\begin{figure}
\centering
\includegraphics[width=0.6\textwidth]{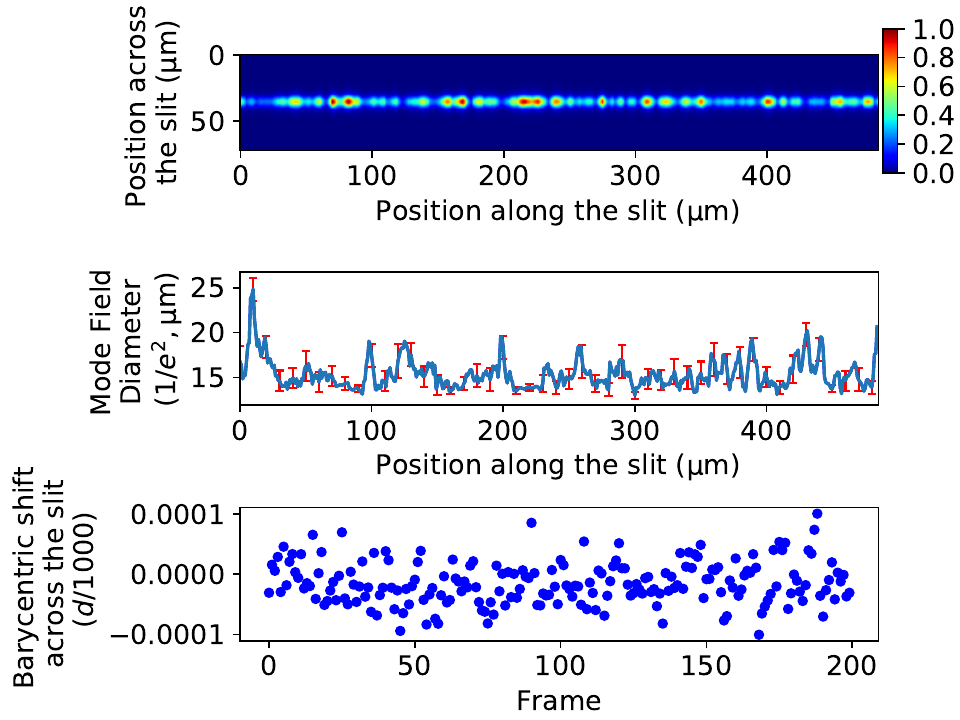}
\caption{Top: Averaged (intensity) near-field
images of the slit output from BeamProp simulations
(@1550 nm). Middle: \ac{MFD} profile size
measurements of the slit with 1$\upsigma$ errors 
from the individual images. Bottom: Barycentric
variations across the slit of the device calculated
from our dataset.}
\label{slit_com}
\end{figure}

The dimension of the \ac{MFD} and its barycentric
movements were measured, from the simulation results.
We also examined whether there are disturbances
of the injected field to the device, which lead to
different speckle distribution at the reformatter's 
slit output. Results of our analysis are presented
in Figure \ref{slit_com}. There the averaged 
(intensity) slit images are shown, as well as the 
calculated \ac{MFD} size from fitting Gaussian 
distributions along the slit, and finally the
barycentric movement of the \ac{MFD} across the
slit. The results of the barycentric movement
of the slit, are expressed as a ratio of one-thousandth
of the waveguide's core diameter size ($d$/1000). It
is immediately apparent from the figure that the 
average semi-amplitude variation of the barycenter 
is of the order of $10^{-4}$ ($d$/1000) which is small
enough and meets our expectations for the modal noise
properties of this device compared to other devices
(Ref. \citenum{Anagnos:2018}), as discussed
further in section \ref{mnoise:dis}. The aforementioned
simulation results were obtained without taking into
account any manufacturing errors in the slit 
straightness. As expected, the existence of such
variations affect negatively the spectral resolving
power of a spectrograph by causing noise and 
uncertainties in the measured spectra.

\section{Discussion}
\label{sec:discussion}

\subsection{Throughput performance}
\label{th_mismatch}

Following the procedure as described in section
\ref{sec:methods} for our simulations, we obtained
a throughput difference between the simulations
and the on-sky output performance, of the order of 
3 to 8 per-cent (see Table \ref{table:res}). 
These small differences in throughput results are
due to the following reasons: first the on-sky
\ac{AO} raw data during the night of observation were 
not available at this point, therefore the \ac{AO}
performance per mode of correction could not be
matched with precision; second the changing on-sky
atmosphere conditions during the observation run;
last, the assumption of perfect waveguide's 
structure and geometry in our simulations, without 
accounting for the manufacturing errors.

\begin{table}[ht]
\centering
\caption{Percentage results of throughput from
simulations in comparison with the on-sky conditions.}
\label{table:res}
\begin{tabular}{lcc}
\hline
&Data and results&\\
\hline \hline
\ac{AO} mode & On-sky & Soapy + BeamProp\\
\hline
closed-loop (\%) & 53 $\pm$ 4 & 56 $\pm$ 4\\
tip-tilt (\%) & 47 $\pm$ 5 & 55 $\pm$ 5\\
open-loop (\%) & 48 $\pm$ 5 & 51 $\pm$ 7\\
\hline
\end{tabular}
\end{table}

\subsection{Coupling of evanescent field}
\label{sec:ev_coupling}

The combination of a \ac{PL} feeding a \ac{ULI}
reformatter creating the \ac{HR} device resulted
in the elimination of the undesired coupling of the
evanescent field into the device, encountered in 
Ref. \citenum{Anagnos:2018}. This is due to the
high refractive index difference among the cores and 
the cladding of the \ac{PL} that guide light only
inside the cores and not in the surrounding cladding.

\subsection{Modal noise}
\label{mnoise:dis}
It is apparent from the bottom part of Figure
\ref{slit_com} that the barycenter movement of the
slit is not completely stable, but has some variations
of minor importance, even though the slit is designed
to be completely straight. As a result, it will 
slightly affect the spectral resolving power of the
spectrograph, but not to the same magnitude as when
compared to a conventional fiber\cite{Chen:2006}.
To confirm our hypothesis, we performed two 
experiments; we applied the same method as in
section \ref{sec:mnoise} to a common state-of-the-art
\ac{MM} fiber with an octagonal cross section of
50 $\upmu$m in diameter \cite{Chen:2006} and a 
common \ac{SM} fiber 8.2 $\upmu$m in diameter
(Corning SMF-28 Optical Fiber).

\begin{figure}
\begin{tabular}{ll@{}}
\includegraphics[width=0.5\textwidth]{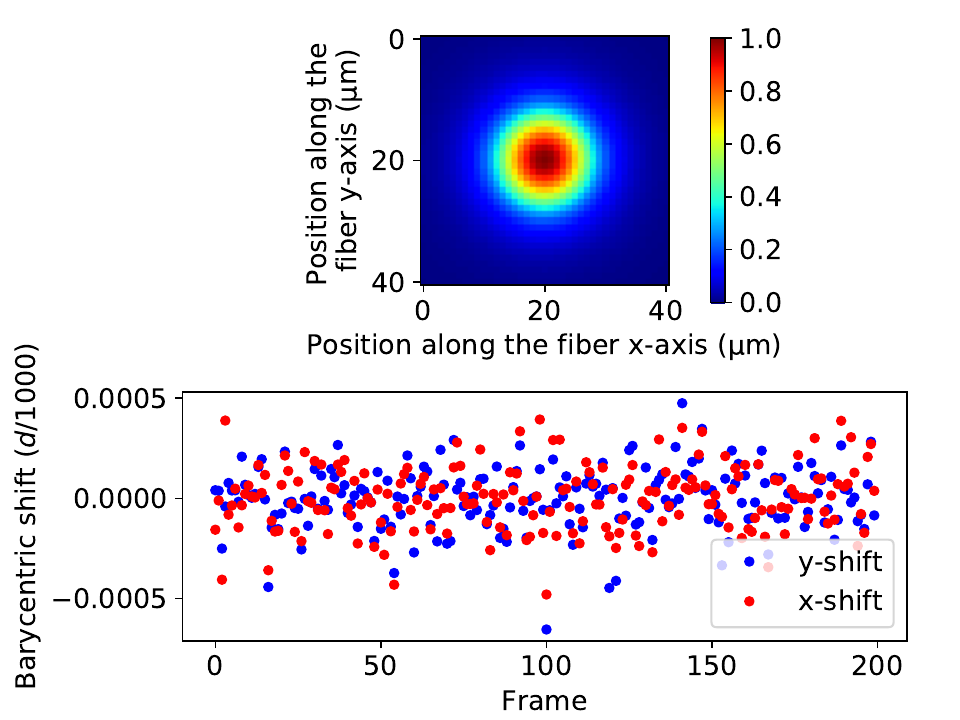}
& \includegraphics[width=0.5\textwidth]{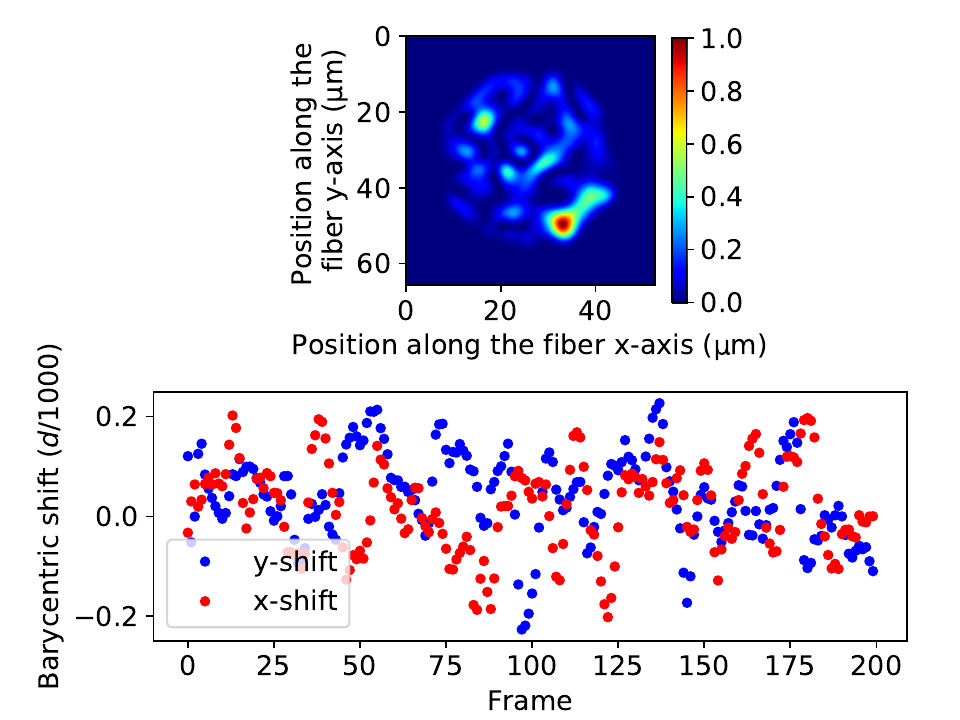}
\end{tabular}
\caption{\textbf{Left sub-figure}: Top: Common
near-field image (intensity) of the 8.2 $\upmu$m 
\ac{SM} fiber output from BeamProp simulations. 
Bottom: Barycentric movement measurements along
the fiber x-axis (red) and y-axis (blue) derived from
individual images. \textbf{Right sub-figure}: Top:
Common near-field image (intensity) of the 50 $\upmu$m
octagonal fiber output from BeamProp simulations.
Bottom: Barycentric movement measurements along
the fiber x-axis (red) and y-axis (blue) derived from
individual images.}
\label{slit:sep}
\end{figure}

It is apparent from results that the \ac{HR} modal
noise properties are similar to the \ac{SM} fiber,
while their barycentric movements are three orders
of magnitude less than the \ac{MM} octagonal fiber
results (see Figure \ref{slit:sep}). Our results 
are quantitatively similar to other related results
in the literature (e.g. Ref. \citenum{Feger:2012}).

\section{Conclusions}
\label{sec:conclusion}

We have performed a simulation experiment regarding
the performance of an existing astrophotonic spatial
reformatter, the \acl{HR}. We used Soapy\cite{Reeves:2016},
a simulation program to replicate the on-sky conditions
and its influence on the device's performance, and
BeamProp by RSoft\cite{rsoft}, a finite-difference
beam propagation solver to model the light propagation
through out device using data produced by Soapy.

Our simulation results were close to the on-sky
reported values, yielding a device performance
of 56 $\pm$ 4\% in closed-loop (compared to 53
$\pm$ 4\% for on-sky), 55 $\pm$ 5\% in tip-tilt
(compared to 47 $\pm$ 5\% for on-sky) and 51 $\pm$ 7\%
in open-loop (compared to 48 $\pm$ 5\% for on-sky).
The variable atmospheric seeing conditions encountered
on-sky during the night of observations, which were not
taken into account in this simulation, explain these
differences in throughput.

Furthermore, we investigated the modal noise 
properties of the \acl{HR}. Our results show
that although the modal noise is not entirely
absent, it is improved by three orders of 
magnitude compared to a typical \ac{MM} fiber
with an octagonal cross section.

Additional simulations were performed to optimize
the device's transmission performance, resulting 
in a 2\% improvement in theoretical transmission
(a $\sim$ 6\% improvement in performance) over 
the model of the original device. Simulating the
performance of components including atmospheric
effects proves to be crucial for improving their
characteristics.

Our method as well as its outcome, shows that
performing similar simulations to enhance the designs
of future astronomical components is essential. Increasing their
measurement precision while allowing for better
calibration, and finally making them more compact
in size and be compatible to the new generation of
\acp{ELT}.

Future plans include the further optimization of
the \acl{HR} \cite{MacLachlan:2017} with a high
potential to manufacture the optimized version.

\section*{Acknowledgements}

This work was supported by the Deutsche Forschungsgemeinschaft 
(DFG) through project 326946494, `Novel Astronomical
Instrumentation through photonic Reformatting'. Robert J.
Harris is funded/supported by the Carl-Zeiss-Foundation.
R.R.T sincerely thanks the UK Science and Technology
Facilities Council (STFC) for support through an STFC
Consortium Grant (ST/N000625/1).

This research made use of Astropy, a community-developed
core Python package for Astronomy (Astropy Collaboration, 2013) \cite{astropy:2018},
Numpy \cite{numpy} and Matplotlib \cite{matplotlib}

\bibliography{report} 
\bibliographystyle{spiebib} 

\end{document}